\documentclass[12pt]{article}
\usepackage{amsmath,amssymb}
\usepackage{bm}
\usepackage{mathrsfs}
\usepackage{fourier}
\usepackage{multirow}
\allowdisplaybreaks[4]
\usepackage[usenames]{color}
\newcommand{{\Slashp}}{p\!\!\!\!\!\big/}
\newcommand{{\Slashq}}{q\!\!\!\!\!\big/}
\newcommand{{\Slashk}}{k\!\!\!\!\!\big/}

\begin{document}

\title{Local particle-ghost symmetry}

\author{
Yoshiharu \textsc{Kawamura}\footnote{E-mail: haru@azusa.shinshu-u.ac.jp}\\
{\it Department of Physics, Shinshu University, }\\
{\it Matsumoto 390-8621, Japan}\\
}

\date{
May 20, 2015}

\maketitle
\begin{abstract}
We study the quantization of systems with local particle-ghost symmetries.
The systems contain ordinary particles including gauge bosons
and their counterparts obeying different statistics.
The particle-ghost symmetry is a kind of fermionic symmetry,
different from the space-time supersymmetry and the BRST symmetry.
Subsidiary conditions on states guarantee the unitarity of systems.
\end{abstract}


\section{Introduction}

Graded Lie algebras or Lie superalgebras have been frequently used to formulate theories 
and construct models in particle physics.
Typical examples are supersymmetry (SUSY)~\cite{NS,R,ACS,GS}
and BRST symmetry~\cite{BRS1,BRS2,T}.

The space-time SUSY~\cite{W&Z,W&B} is a symmetry 
between ordinary particles with integer spin
and those with half-integer spin, and the generators called supercharges
are space-time spinors that obey the anti-commutation relations~\cite{SS,BLS}.

The BRST symmetry is a symmetry concerning 
unphysical modes in gauge fields and abnormal fields called
Faddeev-Popov ghost fields~\cite{F&P}.
Though both gauge fields and abnormal fields contain negative norm states, 
theories become unitary on the physical subspace, 
thanks to the BRST invariance~\cite{K&O1,K&O2}.
The BRST and anti-BRST charges are anti-commuting space-time scalars.

Recently, models that contain both ordinary particles with a positive norm
and their counterparts obeying different statistics
have been constructed and those features have been studied~\cite{YK1,YK2,YK3,YK4}.
Models have fermionic symmetries different from the space-time SUSY
and the BRST symmetry.
We refer to this type of novel symmetries as $\lq\lq$particle-ghost symmetries''.

The particle-ghost symmetries have been introduced as global symmetries,
but we do not need to restrict them to the global ones.
Rather, it would be meaningful to examine systems 
with local particle-ghost symmetries from following reasons.
It is known that any global continuous symmetries can be broken down 
by the effect of quantum gravity such as a wormhole~\cite{G&S}.
Then, it is expected that an fundamental theory possesses local symmetries,
and global continuous symmetries can appear as accidental ones in lower-energy scale.
In the system with global particle-ghost symmetries,
the unitarity holds by imposing subsidiary conditions on states by hand.
In contrast, there is a possibility that
the conditions are realized as remnants of local symmetries
in a specific situation.

We study the quantization of systems with local particle-ghost symmetries.
The systems contain ordinary particles including gauge bosons
and their counterparts obeying different statistics.
Subsidiary conditions on states guarantee the unitarity of systems.
The conditions can be originated from constraints 
in case that gauge fields have no dynamical degrees of freedom.

The contents of this paper are as follows.
We construct models with local fermionic symmetries in Sect. II,
and carry out the quantization of the system 
containing scalar and gauge fields in Sect. III.
Section IV is devoted to conclusions and discussions.
In appendix A, we study the system that gauge fields are auxiliary ones.

\section{Systems with local fermionic symmetries}

\subsection{Scalar fields with local fermionic symmetries}

Recently, the system described by the following Lagrangian density
has been studied~\cite{YK1,YK2,YK3,YK4},
\begin{eqnarray}
\mathcal{L}_{\varphi, c_{\varphi}} 
= \partial_{\mu} \varphi^{\dagger} \partial^{\mu} \varphi 
- m^2 \varphi^{\dagger} \varphi
+ \partial_{\mu} c_{\varphi}^{\dagger} \partial^{\mu} c_{\varphi} 
- m^2 c_{\varphi}^{\dagger} c_{\varphi},
\label{L-varphi-c}
\end{eqnarray}
where $\varphi$ is an ordinary complex scalar field 
and $c_{\varphi}$ is the fermionic counterpart obeying the anti-commutation relations.
The system has a global $OSp(2|2)$ symmetry
that consists of $U(1)$ and fermionic symmetries.
The unitarity holds by imposing suitable subsidiary conditions 
relating the conserved charges on states.

Starting from (\ref{L-varphi-c}),
the model with the local $OSp(2|2)$ symmetry is constructed
by introducing gauge fields.
The resultant Lagrangian density is given by
\begin{eqnarray}
&~& \mathcal{L}= \mathcal{L}_{\rm M} + \mathcal{L}_{\rm G},
\nonumber\\
&~& \mathcal{L}_{\rm M}
= \bigl\{(\partial_{\mu} - i g A_{\mu} - i g B_{\mu}) \varphi^{\dagger} 
- g C_{\mu}^{-} c_{\varphi}^{\dagger}\bigr\}
\bigl\{(\partial^{\mu} + i g A^{\mu} + i g B^{\mu}) \varphi
+ g C^{+\mu} c_{\varphi}\bigr\} 
\nonumber\\
&~& ~~~~~~~~~~~ 
+ \bigl\{(\partial_{\mu} - ig A_{\mu} + i g B_{\mu}) c_{\varphi}^{\dagger} 
- g C_{\mu}^{+} \varphi^{\dagger}\bigr\}
\bigl\{(\partial^{\mu} + i g A^{\mu} - i g B^{\mu}) c_{\varphi}
- g C^{-\mu} \varphi\bigr\} 
\nonumber\\
&~& ~~~~~~~~~~~ 
 - m^2 \varphi^{\dagger} \varphi - m^2 c_{\varphi}^{\dagger} c_{\varphi},
\label{L-M}\\
\hspace{-1cm}&~& \mathcal{L}_{\rm G}
= - \bigl\{\partial_{\mu} A_{\nu} - \partial_{\nu} A_{\mu} 
+ i g (C_{\mu}^{+} C_{\nu}^{-} - C_{\nu}^{+} C_{\mu}^{-})\bigr\}
\bigl\{\partial^{\mu} B^{\nu} - \partial^{\nu} B^{\mu}\bigr\}
\nonumber\\
&~&  ~~~~~~~~~~~ 
- \frac{1}{2} \bigl\{\partial_{\mu} C_{\nu}^{+} - \partial_{\nu} C_{\mu}^{+}
 + 2 i g (B_{\mu} C_{\nu}^+ - B_{\nu} C_{\mu}^+)\bigr\}
\nonumber\\
&~&  ~~~~~~~~~~~~~~~~~~~ \cdot
\bigl\{\partial^{\mu} C^{-\nu} - \partial^{\nu} C^{-\mu}
 - 2 i g (B^{\mu} C^{-\nu} - B^{\nu} C^{-\mu})\bigr\},
\label{L-G}
\end{eqnarray}
where $A_{\mu}$ and $B_{\mu}$ are the gauge fields 
relating the (diagonal) $U(1)$ symmetries, 
$C_{\mu}^{+}$ and $C_{\mu}^{-}$ are gauge fields 
relating the fermionic symmetries, and
$g$ is the gauge coupling constant.
The quantized fields of $C_{\mu}^{\pm}$ obey the anti-commutation relations.

The $\mathcal{L}$ is invariant 
under the local $U(1)$ transformations,
\begin{eqnarray}
&~& \delta_{A} \varphi = - i \epsilon \varphi,~~
\delta_{A} \varphi^{\dagger} = i \epsilon \varphi^{\dagger},~~
\delta_{A} c_{\varphi} =  - i \epsilon c_{\varphi},~~
\delta_{A} c_{\varphi}^{\dagger} = i \epsilon c_{\varphi}^{\dagger},
\nonumber \\
&~& \delta_{A} A_{\mu} = \frac{1}{g} \partial_{\mu} \epsilon,~~
\delta_{A} B_{\mu} = 0,~~
\delta_{A} C_{\mu}^{+} = 0,~~
\delta_{A} C_{\mu}^{-} = 0,
\label{delta-A-local}\\
&~& \delta_{B} \varphi = - i \xi \varphi,~~
\delta_{B} \varphi^{\dagger} = i \xi \varphi^{\dagger},~~
\delta_{B} c_{\varphi} =  i \xi c_{\varphi},~~
\delta_{B} c_{\varphi}^{\dagger} = - i \xi c_{\varphi}^{\dagger},
\nonumber \\
&~& \delta_{B} A_{\mu} = 0,~~
\delta_{B} B_{\mu} = \frac{1}{g} \partial_{\mu} \xi,~~
\delta_{B} C_{\mu}^{+} = - 2 i \xi C_{\mu}^+,~~
\delta_{B} C_{\mu}^{-} = 2 i \xi C_{\mu}^-
\label{delta-B-local}
\end{eqnarray}
and the local fermionic transformations,
\begin{eqnarray}
&~& \delta_{\rm F} \varphi = - \zeta c_{\varphi},~~\delta_{\rm F} \varphi^{\dagger} = 0,~~ 
\delta_{\rm F} c_{\varphi} = 0,~~\delta_{\rm F} c_{\varphi}^{\dagger} = \zeta \varphi^{\dagger},~~
\nonumber \\
&~& \delta_{\rm F} A_{\mu} = - i \zeta C_{\mu}^{-},~~
\delta_{\rm F} B_{\mu} = 0,~~
\delta_{\rm F} C_{\mu}^{+} = 2 i \zeta B_{\mu} + \frac{1}{g} \partial_{\mu} \zeta,~~
\delta_{\rm F} C_{\mu}^{-} =0,~~
\label{delta-F-local}\\
&~& \delta_{\rm F}^{\dagger} \varphi = 0,~~
\delta_{\rm F}^{\dagger} \varphi^{\dagger} = \zeta^{\dagger} c_{\varphi}^{\dagger},~~
\delta_{\rm F}^{\dagger} c_{\varphi} = \zeta^{\dagger} \varphi,~~
\delta_{\rm F}^{\dagger} c_{\varphi}^{\dagger} = 0,~~
\nonumber \\
&~& \delta_{\rm F}^{\dagger} A_{\mu} = - i \zeta^{\dagger} C_{\mu}^{+},~~
\delta_{\rm F}^{\dagger} B_{\mu} = 0,~~
\delta_{\rm F}^{\dagger} C_{\mu}^{+} = 0,~~
\delta_{\rm F}^{\dagger} C_{\mu}^{-} = - 2 i \zeta^{\dagger} B_{\mu}
+ \frac{1}{g} \partial_{\mu} \zeta^{\dagger},~~
\label{delta-Fdagger-local}
\end{eqnarray} 
where $\epsilon$ and $\xi$ are infinitesimal real functions of $x$,
and $\zeta$ and $\zeta^{\dagger}$ 
are Grassmann-valued functions of $x$.

The $\mathcal{L}_{\rm M}$ and $\mathcal{L}_{\rm G}$ are simply written as
\begin{eqnarray}
\mathcal{L}_{\rm M} = (D_{\mu} \Phi)^{\dagger} (D^{\mu} \Phi) 
- m^2 \Phi^{\dagger} \Phi~~~~{\rm and}~~~~
\mathcal{L}_{\rm G}
= -\frac{1}{4} {\rm Str} (F_{\mu\nu}F^{\mu\nu}),
\label{L-again}
\end{eqnarray}
respectively.
In $\mathcal{L}_{\rm M}$, $D_{\mu}$ and $\Phi$ are the covariant derivative
and the doublet of fermionic transformation defined by
\begin{eqnarray}
D_{\mu} \equiv
\left(
\begin{array}{cc}
\partial_{\mu} + i g A_{\mu} + i g B_{\mu} & g C_{\mu}^{+} \\
- g C_{\mu}^{-} & \partial_{\mu} + i g A_{\mu} - i g B_{\mu}  
\end{array}
\right)~~~~{\rm and}~~~~
\Phi \equiv 
\left(
\begin{array}{c}
\varphi \\
c_{\varphi}
\end{array}
\right),
\label{DmuPhi}
\end{eqnarray}
respectively.
In $\mathcal{L}_{\rm G}$,
${\rm Str}$ is the supertrace defined by ${\rm Str}M = a - d$
where $M$ is the $2 \times 2$ matrix given by
\begin{eqnarray}
M =
\left(
\begin{array}{cc}
a & b \\
c & d 
\end{array}
\right).
\label{M}
\end{eqnarray}
The $F_{\mu\nu}$ is defined by
\begin{eqnarray}
F_{\mu\nu} \equiv \frac{1}{i g} [D_{\mu}, D_{\nu}] =
\left(
\begin{array}{cc}
A_{\mu\nu} + B_{\mu\nu} & - i C_{\mu\nu}^{+} \\
i C_{\mu\nu}^{-} & A_{\mu\nu} - B_{\mu\nu}  
\end{array}
\right),
\label{Fmunu}
\end{eqnarray}
where $A_{\mu\nu}$, $B_{\mu\nu}$, $C_{\mu\nu}^{+}$ and $C_{\mu\nu}^{-}$ 
are the field strengths given by
\begin{eqnarray}
&~& A_{\mu\nu} = \partial_{\mu} A_{\nu} - \partial_{\nu} A_{\mu}
+ i g (C_{\mu}^{+} C_{\nu}^{-} - C_{\nu}^{+} C_{\mu}^{-}),
\label{Amunu}\\
&~& B_{\mu\nu} = \partial_{\mu} B_{\nu} - \partial_{\nu} B_{\mu},
\label{Bmunu}\\
&~& C_{\mu\nu}^{+} = \partial_{\mu} C_{\nu}^{+} - \partial_{\nu} C_{\mu}^{+}
+ 2 i g (B_{\mu} C_{\nu}^+ - B_{\nu} C_{\mu}^+),~~ 
\label{C+munu}\\
&~& C_{\mu\nu}^{-} = \partial_{\mu} C_{\nu}^{-} - \partial_{\nu} C_{\mu}^{-}
- 2 i g (B_{\mu} C_{\nu}^{-} - B_{\nu} C_{\mu}^{-}).
\label{C-munu}
\end{eqnarray}
Under the transformations (\ref{delta-A-local}) -- (\ref{delta-Fdagger-local}), 
the field strengths are transformed as
\begin{eqnarray}
&~& \delta_{A} A_{\mu\nu} = 0,~~\delta_{A} B_{\mu\nu} = 0,~~
\delta_{A} C_{\mu\nu}^{+} = 0,~~  \delta_{A} C_{\mu\nu}^{-} = 0,
\label{fs-A}\\
&~& \delta_{B} A_{\mu\nu} = 0,~~\delta_{B} B_{\mu\nu} = 0,~~
\delta_{B} C_{\mu\nu}^{+} = - 2 i \xi C_{\mu\nu}^{+},~~ 
\delta_{B} C_{\mu\nu}^{-} = 2 i \xi C_{\mu\nu}^{-},
\label{fs-B}\\
&~& \delta_{\rm F} A_{\mu\nu} = - i \zeta C_{\mu\nu}^{-},~~
\delta_{\rm F} B_{\mu\nu} = 0,~~
\delta_{\rm F} C_{\mu\nu}^{+} =  2 i \zeta B_{\mu\nu},~~  \delta_{\rm F} C_{\mu\nu}^{-} = 0,
\label{fs-F}\\
&~& \delta_{\rm F}^{\dagger} A_{\mu\nu} = -i \zeta^{\dagger} C_{\mu\nu}^{+},~~
\delta_{\rm F}^{\dagger} B_{\mu\nu} = 0,~~
\delta_{\rm F}^{\dagger} C_{\mu\nu}^{+} = 0,~~  
\delta_{\rm F}^{\dagger} C_{\mu\nu}^{-} = - 2 i \zeta^{\dagger} B_{\mu\nu}.
\label{fs-F-dagger}
\end{eqnarray}

Using the global fermionic transformations,
\begin{eqnarray}
&~& \tilde{\bm{\delta}}_{\rm F} \varphi 
= - c_{\varphi},~~\tilde{\bm{\delta}}_{\rm F}\varphi^{\dagger} = 0,~~ 
\tilde{\bm{\delta}}_{\rm F} c_{\varphi} = 0,~~
\tilde{\bm{\delta}}_{\rm F} c_{\varphi}^{\dagger} = \varphi^{\dagger},~~
\nonumber \\
&~& \tilde{\bm{\delta}}_{\rm F} A_{\mu} = - i C_{\mu}^{-},~~
\tilde{\bm{\delta}}_{\rm F} B_{\mu} = 0,~~
\tilde{\bm{\delta}}_{\rm F} C_{\mu}^{+} = 2 i B_{\mu},~~
\tilde{\bm{\delta}}_{\rm F} C_{\mu}^{-} = 0,~~
\label{delta-F}\\
&~& \tilde{\bm{\delta}}^{\dagger}_{\rm F} \varphi = 0,~~
\tilde{\bm{\delta}}^{\dagger}_{\rm F} \varphi^{\dagger} = c_{\varphi}^{\dagger},~~
\tilde{\bm{\delta}}^{\dagger}_{\rm F} c_{\varphi} = \varphi,~~
\tilde{\bm{\delta}}^{\dagger}_{\rm F} c_{\varphi}^{\dagger} = 0,~~
\nonumber \\
&~& \tilde{\bm{\delta}}^{\dagger}_{\rm F} A_{\mu} = - i C_{\mu}^{+},~~
\tilde{\bm{\delta}}^{\dagger}_{\rm F} B_{\mu} = 0,~~
\tilde{\bm{\delta}}^{\dagger}_{\rm F} C_{\mu}^{+} = 0,~~
\tilde{\bm{\delta}}^{\dagger}_{\rm F} C_{\mu}^{-} = - 2 i B_{\mu},~~
\label{delta-Fdagger}
\end{eqnarray}
$\mathcal{L}$ is rewritten as
\begin{eqnarray}
\mathcal{L}= \tilde{\bm{\delta}}_{\rm F} \tilde{\bm{\delta}}^{\dagger}_{\rm F} 
\mathcal{L}_{\varphi, A}
= - \tilde{\bm{\delta}}^{\dagger}_{\rm F} \tilde{\bm{\delta}}_{\rm F} 
\mathcal{L}_{\varphi, A},
\label{L-exact}
\end{eqnarray}
where $\mathcal{L}_{\varphi, A}$ is given by
\begin{eqnarray}
&~& \mathcal{L}_{\varphi, A}
= \bigl\{(\partial_{\mu} - i g A_{\mu} - i g B_{\mu}) \varphi^{\dagger} 
- g C_{\mu}^{-} c_{\varphi}^{\dagger}\bigr\}
\bigl\{(\partial^{\mu} + i g A^{\mu} + i g B^{\mu}) \varphi
+ g C^{+\mu} c_{\varphi}\bigr\} 
\nonumber \\
&~& ~~~~~~~~~~~~~~
 - m^2 \varphi^{\dagger} \varphi
- \frac{1}{4} A_{\mu\nu} A^{\mu\nu}.
\label{L-varphiA}
\end{eqnarray}

\subsection{Spinor fields with local fermionic symmetries}

For spinor fields, we consider the Lagrangian density,
\begin{eqnarray}
\mathcal{L}_{\psi, c_{\psi}} 
= i \overline{\psi} \gamma^{\mu} \partial_{\mu} \psi - m \overline{\psi} \psi
+ i \overline{c}_{\psi} \gamma^{\mu} \partial_{\mu} c_{\psi} - m \overline{c}_{\psi}c_{\psi},
\label{L-psi-c}
\end{eqnarray}
where $\psi$ is an ordinary spinor field and 
$c_{\psi}$ is its bosonic counterpart obeying commutation relations.
This system also has global $U(1)$ and fermionic symmetries,
and the unitarity holds by imposing suitable subsidiary conditions on states.

Starting from (\ref{L-psi-c}),
the Lagrangian density with local symmetries is constructed as
\begin{eqnarray}
&~& \mathcal{L}^{\rm sp}= \mathcal{L}_{\rm M}^{\rm sp} + \mathcal{L}_{\rm G},
\nonumber\\
&~& \mathcal{L}_{\rm M}^{\rm sp}
=  i \overline{\psi} \gamma^{\mu} 
\bigl\{(\partial_{\mu} + i g A_{\mu} + i g B_{\mu}) \psi
+ g C_{\mu}^{+} c_{\psi}\bigr\}  - m \overline{\psi} \psi
\nonumber\\
&~& ~~~~~~~~~~~ +  i \overline{c}_{\psi} \gamma^{\mu}
\bigl\{(\partial_{\mu} + i g A_{\mu} - i g B_{\mu}) c_{\psi} - g C_{\mu}^{-} \psi\bigr\} 
 - m \overline{c}_{\psi} c_{\psi},
\label{L-M-sp}
\end{eqnarray}
where $\mathcal{L}_{\rm G}$ is given by (\ref{L-G}),
$\overline{\psi} \equiv \psi^{\dagger} \gamma^0$,
$\overline{c}_{\psi} \equiv c_{\psi}^{\dagger} \gamma^0$
and $\gamma^{\mu}$ are the $\gamma$ matrices satisfying 
$\{\gamma^{\mu}, \gamma^{\nu}\} = 2 \eta^{\mu\nu}$.
The $\mathcal{L}_{\rm M}^{\rm sp}$ is rewritten as
\begin{eqnarray}
 \mathcal{L}_{\rm M}^{\rm sp}
= i \overline{\Psi} \Gamma^{\mu} D_{\mu} \Psi - m \overline{\Psi} \Psi,
\label{L-M-sp2}
\end{eqnarray}
where $\Gamma^{\mu}$ and $\Psi$ are
the extension of $\gamma$-matrices
and the doublet of fermionic transformation defined by
\begin{eqnarray}
\Gamma^{\mu} \equiv
\left(
\begin{array}{cc}
\gamma^{\mu} & 0 \\
0 & \gamma^{\mu}
\end{array}
\right)~~~~ {\rm and}~~~~
\Psi \equiv 
\left(
\begin{array}{c}
\psi \\
c_{\psi}
\end{array}
\right),
\label{DmuPsi}
\end{eqnarray}
respectively.

The $\mathcal{L}^{\rm sp}$ is invariant 
under the local $U(1)$ transformations,
\begin{eqnarray}
&~& \delta_{A} \psi = - i \epsilon \psi,~~
\delta_{A} \psi^{\dagger} = i \epsilon \psi^{\dagger},~~
\delta_{A} c_{\psi} =  - i \epsilon c_{\psi},~~
\delta_{A} c_{\psi}^{\dagger} = i \epsilon c_{\psi}^{\dagger},
\nonumber \\
&~& \delta_{A} A_{\mu} = \frac{1}{g} \partial_{\mu} \epsilon,~~
\delta_{A} B_{\mu} = 0,~~
\delta_{A} C_{\mu}^{+} = 0,~~
\delta_{A} C_{\mu}^{-} = 0,
\label{delta-A-local-sp}\\
&~& \delta_{B} \psi = - i \xi \psi,~~
\delta_{B} \psi^{\dagger} = i \xi \psi^{\dagger},~~
\delta_{B} c_{\psi} =  i \xi c_{\psi},~~
\delta_{B} c_{\psi}^{\dagger} = -i \xi c_{\psi}^{\dagger},
\nonumber \\
&~& \delta_{B} A_{\mu} = 0,~~
\delta_{B} B_{\mu} = \frac{1}{g} \partial_{\mu} \xi,~~
\delta_{B} C_{\mu}^{+} = - 2 i \xi C_{\mu}^+,~~
\delta_{B} C_{\mu}^{-} = 2 i \xi C_{\mu}^-
\label{delta-B-local-sp}
\end{eqnarray}
and the local fermionic transformations,
\begin{eqnarray}
&~& \delta_{\rm F} \psi = - \zeta c_{\psi},~~\delta_{\rm F} \psi^{\dagger} = 0,~~ 
\delta_{\rm F} c_{\psi} = 0,~~\delta_{\rm F} c_{\psi}^{\dagger} = - \zeta \psi^{\dagger},~~
\nonumber \\
&~& \delta_{\rm F} A_{\mu} = - i \zeta C_{\mu}^{-},~~
\delta_{\rm F} B_{\mu} = 0,~~
\delta_{\rm F} C_{\mu}^{+} = 2 i \zeta B_{\mu} + \frac{1}{g} \partial_{\mu} \zeta,~~
\delta_{\rm F} C_{\mu}^{-} =0,~~
\label{delta-F-local-sp}\\
&~& \delta_{\rm F}^{\dagger} \psi = 0,~~
\delta_{\rm F}^{\dagger} \psi^{\dagger} = - \zeta^{\dagger} c_{\psi}^{\dagger},~~
\delta_{\rm F}^{\dagger} c_{\psi} = \zeta^{\dagger} \psi,~~
\delta_{\rm F}^{\dagger} c_{\psi}^{\dagger} = 0,~~
\nonumber \\
&~& \delta_{\rm F}^{\dagger} A_{\mu} = - i \zeta^{\dagger} C_{\mu}^{+},~~
\delta_{\rm F}^{\dagger} B_{\mu} = 0,~~
\delta_{\rm F}^{\dagger} C_{\mu}^{+} = 0,~~
\delta_{\rm F}^{\dagger} C_{\mu}^{-} = - 2 i \zeta^{\dagger} B_{\mu}
+ \frac{1}{g} \partial_{\mu} \zeta^{\dagger},~~
\label{delta-Fdagger-local-sp}
\end{eqnarray} 
where $\epsilon$ and $\xi$ are infinitesimal real functions of $x$,
and $\zeta$ and $\zeta^{\dagger}$ 
are Grassmann-valued functions of $x$.

Using the global fermionic transformations,
\begin{eqnarray}
&~& \tilde{\bm{\delta}}_{\rm F} \psi = - c_{\psi},~~
\tilde{\bm{\delta}}_{\rm F} \psi^{\dagger} = 0,~~ 
\tilde{\bm{\delta}}_{\rm F} c_{\psi} = 0,~~
\tilde{\bm{\delta}}_{\rm F} c_{\psi}^{\dagger} = - \psi^{\dagger},~~
\nonumber \\
&~& \tilde{\bm{\delta}}_{\rm F} A_{\mu} = - i C_{\mu}^{-},~~
\tilde{\bm{\delta}}_{\rm F} B_{\mu} = 0,~~
\tilde{\bm{\delta}}_{\rm F} C_{\mu}^{+} = 2 i B_{\mu},~~
\tilde{\bm{\delta}}_{\rm F} C_{\mu}^{-} = 0,~~
\label{delta-F-sp}\\
&~& \tilde{\bm{\delta}}^{\dagger}_{\rm F} \psi = 0,~~
\tilde{\bm{\delta}}^{\dagger}_{\rm F} \psi^{\dagger} = - c_{\psi}^{\dagger},~~
\tilde{\bm{\delta}}^{\dagger}_{\rm F} c_{\psi} = \psi,~~
\tilde{\bm{\delta}}^{\dagger}_{\rm F} c_{\psi}^{\dagger} = 0,~~
\nonumber \\
&~& \tilde{\bm{\delta}}^{\dagger}_{\rm F} A_{\mu} = - i C_{\mu}^{+},~~
\tilde{\bm{\delta}}^{\dagger}_{\rm F} B_{\mu} = 0,~~
\tilde{\bm{\delta}}^{\dagger}_{\rm F} C_{\mu}^{+} = 0,~~
\tilde{\bm{\delta}}^{\dagger}_{\rm F} C_{\mu}^{-} = - 2 i B_{\mu},~~
\label{delta-Fdagger-sp}
\end{eqnarray}
$\mathcal{L}^{\rm sp}$ is rewritten as
\begin{eqnarray}
\mathcal{L}= \tilde{\bm{\delta}}_{\rm F} \tilde{\bm{\delta}}^{\dagger}_{\rm F} 
\mathcal{L}_{\psi, A} 
= - \tilde{\bm{\delta}}^{\dagger}_{\rm F} \tilde{\bm{\delta}}_{\rm F} 
\mathcal{L}_{\psi, A},
\label{L-exact-sp}
\end{eqnarray}
where $\mathcal{L}_{\psi, A}$ is given by
\begin{eqnarray}
&~& \mathcal{L}_{\psi, A}
=  i \overline{\psi} \gamma^{\mu} 
\bigl\{(\partial_{\mu} + i g q A_{\mu} + i g B_{\mu}) \psi
+ g C_{\mu}^{+} c_{\psi}\bigr\}  - m \overline{\psi} \psi
- \frac{1}{4} A_{\mu\nu} A^{\mu\nu}.
\label{L-psiA}
\end{eqnarray}

\section{Quantization}

We carry out the quantization of the system 
with scalar and gauge fields
described by $\mathcal{L} = \mathcal{L}_{\rm M} + \mathcal{L}_{\rm G}$.

\subsection{Canonical quantization}

Based on the formulation with the property that
{\it the hermitian conjugate of canonical momentum for a variable is 
just the canonical momentum for the hermitian conjugate of the variable}~\cite{YK2}, 
the conjugate momenta are given by
\begin{eqnarray}
&~& \pi \equiv \left(\frac{\partial \mathcal{L}}{\partial \dot{\varphi}}\right)_{\rm R}
=  (\partial_{0} - ig A_0 - i g B_{0} ) {\varphi}^{\dagger} - g C_{0}^{-} c_{\varphi}^{\dagger},~~
\label{pi}\\
&~& \pi^{\dagger} \equiv 
\left(\frac{\partial \mathcal{L}}{\partial \dot{\varphi}^{\dagger}}\right)_{\rm L}
= (\partial_{0} + ig A_0 + i g B_{0}) {\varphi} + g C_{0}^{+} c_{\varphi},
\label{pi-dagger}\\
&~& \pi_{c_{\varphi}} \equiv 
\left(\frac{\partial \mathcal{L}}{\partial \dot{c}_{\varphi}}\right)_{\rm R}
=  (\partial_{0} - ig A_0 + i g B_{0}) c_{\varphi}^{\dagger}  - g C_{0}^{+} \varphi^{\dagger},~~
\label{pi-c}\\
&~& \pi_{c_{\varphi}}^{\dagger} \equiv 
\left(\frac{\partial \mathcal{L}}{\partial \dot{c}_{\varphi}^{\dagger}}\right)_{\rm L}
=  (\partial_{0} + ig A_0 - i g B_{0}) c_{\varphi}  - g C_{0}^{-} \varphi,~~
\label{pi-c-dagger}\\
&~& \Pi_{A}^{\mu} \equiv
\left(\frac{\partial \mathcal{L}}{\partial \dot{A}_{\mu}}\right)_{\rm L} = 2 B^{\mu 0},~~
\Pi_{B}^{\mu} \equiv
\left(\frac{\partial \mathcal{L}}{\partial \dot{B}_{\mu}}\right)_{\rm R} = 2 A^{\mu 0},~~
\label{Pi-AB}\\
&~& \Pi_{C}^{+\mu} \equiv
\left(\frac{\partial \mathcal{L}}{\partial \dot{C}_{\mu}^{+}}\right)_{\rm L} 
= C^{-\mu 0},~~
\Pi_{C}^{-\mu} \equiv
\left(\frac{\partial \mathcal{L}}{\partial \dot{C}_{\mu}^{-}}\right)_{\rm R} 
= C^{+\mu 0},~~
\label{Pi-C}
\end{eqnarray}
where $\dot{\mathcal{O}} = \partial \mathcal{O}/\partial t$,
and R and L stand for the right-differentiation and the left-differentiation, respectively.
From (\ref{Pi-AB}) and (\ref{Pi-C}),
we obtain the primary constraints,
\begin{eqnarray}
\Pi_{A}^{0}  = 0,~~
\Pi_{B}^{0}  = 0,~~
\Pi_{C}^{+0} = 0,~~
\Pi_{C}^{-0} = 0.
\label{primary}
\end{eqnarray}

Using the Legendre transformation,
the Hamiltonian density is obtained as
\begin{eqnarray}
&~& \mathcal{H} = \pi \dot{\varphi} 
+ \dot{\varphi}^{\dagger}\pi^{\dagger} + \pi_{c_{\varphi}} \dot{c}_{\varphi} 
+ \dot{c}_{\varphi}^{\dagger} \pi_{c_{\varphi}}^{\dagger}
+ \dot{A}_{\mu} \Pi_{A}^{\mu}  + \Pi_{B}^{\mu} \dot{B}_{\mu}
+ \dot{C}_{\mu}^{+} \Pi_{C}^{+\mu} + \Pi_{C}^{-\mu} \dot{C}_{\mu}^{-}
- \mathcal{L} 
\nonumber \\
&~& ~~~~~~~~~ + \lambda_{A} \Pi_{A}^{0} + \Pi_{B}^{0} \lambda_{B} 
+ \lambda_{C}^{+} \Pi_{C}^{+0} + \Pi_{C}^{-0} \lambda_{C}^{-}
\nonumber \\
&~& ~~~~~~ = \pi \pi^{\dagger} + \pi_{c_{\varphi}} \pi_{c_{\varphi}}^{\dagger} 
+ (D_i \Phi)^{\dagger} (D^i \Phi) + m^2 \Phi^{\dagger} \Phi
\nonumber \\
&~& ~~~~~~~~~ - i g A_0 (\pi \varphi - \varphi^{\dagger} \pi^{\dagger} 
+ \pi_{c_{\varphi}} c_{\varphi} - c_{\varphi}^{\dagger} \pi_{c_{\varphi}}^{\dagger})
\nonumber \\
&~& ~~~~~~~~~ - i g B_0 (\pi \varphi - \varphi^{\dagger} \pi^{\dagger} 
- \pi_{c_{\varphi}} c_{\varphi} + c_{\varphi}^{\dagger} \pi_{c_{\varphi}}^{\dagger}
+ 2 C_i^{+} \Pi_{C}^{+i} - 2 \Pi_{C}^{-i} C_i^{-})
\nonumber \\
&~& ~~~~~~~~~ - g C_0^{+} (\pi c_{\varphi} - \varphi^{\dagger} \pi_{c_{\varphi}}^{\dagger}
+ i C_i^{-} \Pi_A^i - 2 i B_i \Pi_C^{+i})
\nonumber \\
&~& ~~~~~~~~~ - g (c_{\varphi}^{\dagger} \pi^{\dagger} - \pi_{c_{\varphi}} \varphi
- i C_i^{+} \Pi_A^i + 2 i B_i \Pi_C^{-i}) C_0^{-}
\nonumber \\
&~& ~~~~~~~~~ + \Pi_{A i} \Pi_{B}^{i} + A_{ij} B^{ij}
+ \partial_i A_0~\Pi_{A}^{i} + \Pi_{B}^{i} \partial_i B_0
\nonumber \\
&~& ~~~~~~~~~ + \Pi_{C i}^{+} \Pi_{C}^{-i} + \frac{1}{2} C_{ij} C^{ij}
+ \partial_i C_0^{+}~\Pi_{C}^{+i} + \Pi_{C}^{-i} \partial_i C_0^{-}
\nonumber \\
&~& ~~~~~~~~~ +  \lambda_{A} \Pi_{A}^{0} + \Pi_{B}^{0} \lambda_{B}
+ \lambda_{C}^{+} \Pi_{C}^{+0} + \Pi_{C}^{-0} \lambda_{C}^{-},
\label{H}
\end{eqnarray}
where Roman indices $i$ and $j$ denote the spatial components and run from 1 to 3,
$\lambda_{A}$, $\lambda_{B}$, $\lambda_{C}^{+}$ and $\lambda_{C}^{-}$
are Lagrange multipliers, and $\dot{A}_{0} + \lambda_{A}$, 
$\dot{B}_{0} + \lambda_{B}$, 
$\dot{C}_{0}^{+} + \lambda_{C}^{+}$ and $\dot{C}_{0}^{-} + \lambda_{C}^{-}$ 
are rewritten as $\lambda_{A}$, $\lambda_{B}$, $\lambda_{C}^{+}$ and $\lambda_{C}^{-}$
in the final expression.

Secondary constraints are obtained as follows,
\begin{eqnarray}
&~& \frac{d\Pi_{A}^{0}}{dt} = \left\{\Pi_{A}^{0}, H\right\}_{\rm PB} 
= i g \left(\pi \varphi - \varphi^{\dagger} \pi^{\dagger}
+ \pi_{c_\varphi} c_{\varphi} 
 - c_{\varphi}^{\dagger} \pi_{c_{\varphi}}^{\dagger}\right)
+ \partial_i \Pi_{A}^{i} = 0,~~
\label{secondaryA}\\
&~& \frac{d \Pi_{B}^{0}}{dt} = \left\{\Pi_{B}^{0}, H\right\}_{\rm PB} 
= i g \left(\pi \varphi - \varphi^{\dagger} \pi^{\dagger}
- \pi_{c_\varphi} c_{\varphi} 
+ c_{\varphi}^{\dagger} \pi_{c_{\varphi}}^{\dagger} \right.
\nonumber \\
&~& ~~~~~~~~~~~~~~~~~~~~~~~~~~~~~~~~~~~~~~~~ \left.
+ 2 C_i^{+} \Pi_{C}^{+i} - 2 \Pi_{C}^{-i} C_i^{-}\right)
+ \partial_i \Pi_{B}^{i} = 0,~~
\label{secondaryB}\\
&~& \frac{d \Pi_{C}^{+0}}{dt} = \left\{\Pi_{C}^{+0}, H\right\}_{\rm PB} 
= g \left(\pi c_{\varphi} - \varphi^{\dagger} \pi_{c_{\varphi}}^{\dagger}
+ i C_i^{-} \Pi_A^i - 2 i B_i \Pi_C^{+i}\right)
+ \partial_i \Pi_C^{+i} = 0,~~
\label{secondaryC+}\\
&~& \frac{d \Pi_{C}^{-0}}{dt} = \left\{\Pi_{C}^{-0}, H\right\}_{\rm PB} 
= g \left(c_{\varphi}^{\dagger} \pi^{\dagger} - \pi_{c_{\varphi}} \varphi
- i C_i^{+} \Pi_A^i + 2 i B_i \Pi_C^{-i}\right)
+ \partial_i \Pi_C^{-i} = 0,
\label{secondaryC-}
\end{eqnarray}
where $H$ is the Hamiltonian $H = \int \mathcal{H} d^3x$ and
$\{A, B\}_{\rm PB}$ is the Poisson bracket.
The Poisson bracket for the system with 
canonical variables $(Q_k, P_k)$ and $(Q_k^{\dagger}, P_k^{\dagger})$
is defined by~\cite{YK2}
\begin{eqnarray}
&~& \left\{f, g\right\}_{\rm PB} \equiv 
\sum_{k} \left[\left(\frac{\partial f}{\partial Q_k}\right)_{\rm R}
 \left(\frac{\partial g}{\partial P_k}\right)_{\rm L}
- (-)^{|Q_k|} \left(\frac{\partial f}{\partial P_k}\right)_{\rm R}
 \left(\frac{\partial g}{\partial Q_k}\right)_{\rm L} \right.
\nonumber \\
&~& ~~~~~~~~~~~~~~~~~~~~~~~
\left. +  (-)^{|Q_k|} \left(\frac{\partial f}{\partial Q_k^{\dagger}}\right)_{\rm R}
\left(\frac{\partial g}{\partial P_k^{\dagger}}\right)_{\rm L}
- \left(\frac{\partial f}{\partial P_k^{\dagger}}\right)_{\rm R}
\left(\frac{\partial g}{\partial Q_k^{\dagger}}\right)_{\rm L}
\right],
\label{Poisson}
\end{eqnarray}
where $|Q_k|$ is the number representing the Grassmann parity of $Q_k$, 
i.e., $|Q_k|=1$ for the Grassmann odd $Q_k$
and $|Q_k|=0$ for the Grassmann even $Q_k$.
There appear no other constraints, and all constraints are first class ones
and generate local transformations.

We take the gauge fixing conditions,
\begin{eqnarray}
A^{0} = 0,~~ B^{0}  = 0,~~ {C}^{+0} = 0,~~ {C}^{-0} = 0,~~
\partial_i A^{i} = 0,~~ \partial_i B^{i} = 0,~~
\partial_i C^{+i} = 0,~~ \partial_i C^{-i} = 0.
\label{gf}
\end{eqnarray}

The system is quantized by regarding variables as operators
and imposing the following relations 
on the canonical pairs,
\begin{eqnarray}
&~& [\varphi(\bm{x}, t), \pi(\bm{y}, t)] = i \delta^3(\bm{x}-\bm{y}),~~ 
[\varphi^{\dagger}(\bm{x}, t), \pi^{\dagger}(\bm{y}, t)] = i \delta^3(\bm{x}-\bm{y}),
\label{CCR-varphi}\\
&~& \{c_{\varphi}(\bm{x}, t), \pi_{c_{\varphi}}(\bm{y}, t)\} = i \delta^3(\bm{x}-\bm{y}),~~ 
\{c_{\varphi}^{\dagger}(\bm{x}, t), \pi_{c_{\varphi}}^{\dagger}(\bm{y}, t)\} = -i \delta^3(\bm{x}-\bm{y}),
\label{CCR-c}\\
&~& [A_{i}(\bm{x}, t), \Pi_{A}^j(\bm{y}, t)] 
= i \left(\delta_{i}^{j} - \frac{\partial_i \partial^j}{\varDelta}\right) \delta^3(\bm{x}-\bm{y}),~~
\label{CCR-A}\\
&~& [B_{i}(\bm{x}, t), \Pi_{B}^j(\bm{y}, t)] 
= i \left(\delta_{i}^{j} - \frac{\partial_i \partial^j}{\varDelta}\right) \delta^3(\bm{x}-\bm{y}),~~
\label{CCR-B}\\
&~& \{C_{i}^{+}(\bm{x}, t), \Pi_{C}^{+j}(\bm{y}, t)\} 
= - i \left(\delta_{i}^{j} - \frac{\partial_i \partial^j}{\varDelta}\right) \delta^3(\bm{x}-\bm{y}),~~
\label{CCR-C+}\\
&~& \{C_{i}^{-}(\bm{x}, t), \Pi_{C}^{-j}(\bm{y}, t)\} 
= i \left(\delta_{i}^{j} - \frac{\partial_i \partial^j}{\varDelta}\right) \delta^3(\bm{x}-\bm{y}),
\label{CCR-C-}
\end{eqnarray}
where $[\mathcal{O}_1, \mathcal{O}_2] 
\equiv \mathcal{O}_1 \mathcal{O}_2 - \mathcal{O}_2 \mathcal{O}_1$,
$\{\mathcal{O}_1, \mathcal{O}_2\} 
\equiv \mathcal{O}_1 \mathcal{O}_2 + \mathcal{O}_2 \mathcal{O}_1$,
and only the non-vanishing ones are denoted.
Here,
we define the Dirac bracket using the first class constraints and the gauge fixing conditions,
and replace the bracket with the commutator or the anti-commutator.

On the reduced phase space,
the conserved $U(1)$ charges $N_{A}$ and $N_{B}$ 
and the conserved fermionic charges $Q_{\rm F}$ and $Q_{\rm F}^{\dagger}$
are constructed  as
\begin{eqnarray}
&~& N_{A} = - i \int d^3x~\left(\pi \varphi - \varphi^{\dagger} \pi^{\dagger}
+ \pi_{c_\varphi} c_{\varphi} 
 - c_{\varphi}^{\dagger} \pi_{c_{\varphi}}^{\dagger}\right),
\label{NA}\\
&~& N_{B} = - i \int d^3x~\left(\pi \varphi - \varphi^{\dagger} \pi^{\dagger}
- \pi_{c_\varphi} c_{\varphi} 
 + c_{\varphi}^{\dagger} \pi_{c_{\varphi}}^{\dagger}
+ 2 C_i^{+} \Pi_{C}^{+i} - 2 \Pi_{C}^{-i} C_i^{-}\right),
\label{NB}\\
&~& Q_{\rm F} = - \int d^3x~\left(\pi c_{\varphi} 
- \varphi^{\dagger} \pi_{c_{\varphi}}^{\dagger}
+ i C_i^{-} \Pi_A^i - 2 i B_i \Pi_C^{+i}\right),~~
\label{QF}\\
&~& Q_{\rm F}^{\dagger} = - \int d^3x~\left(c_{\varphi}^{\dagger} \pi^{\dagger}
- \pi_{c_{\varphi}} \varphi
- i C_i^{+} \Pi_A^i + 2 i B_i \Pi_C^{-i}\right).
\label{QF-dagger}
\end{eqnarray}
The following algebraic relations hold:
\begin{eqnarray}
&~& {Q_{\rm F}}^2 = 0,~~{Q_{\rm F}^{\dagger}}^2 = 0,~~
\{Q_{\rm F}, Q_{\rm F}^{\dagger}\} = N_{A},~~
[N_{A}, Q_{\rm F}] = 0,~~ [N_{A}, Q_{\rm F}^{\dagger}] = 0,~~
\nonumber\\
&~& [N_{B}, Q_{\rm F}] = - 2 Q_{\rm F},~~
[N_{B}, Q_{\rm F}^{\dagger}] = 2 Q_{\rm F}^{\dagger},~~
[N_{A}, N_{B}] = 0.
\label{QQdagger-varphi}
\end{eqnarray}
The above charges are generators of global $U(1)$
and fermionic transformations such that
\begin{eqnarray}
\tilde{\delta}_{A} \mathcal{O} = i[\epsilon_{0} N_{A}, \mathcal{O}],~~
\tilde{\delta}_{B} \mathcal{O} = i[\xi_{0} N_{B}, \mathcal{O}],~~
\tilde{\delta}_{\rm F} \mathcal{O} = i[\zeta_{0} Q_{\rm F}, \mathcal{O}],~~
\tilde{\delta}_{\rm F}^{\dagger} \mathcal{O} 
= i[Q_{\rm F}^{\dagger}\zeta^{\dagger}_{0}, \mathcal{O}],
\label{delta}
\end{eqnarray}
where $\epsilon_{0}$ and $\xi_{0}$ are real parameters, and
$\zeta_{0}$ and $\zeta^{\dagger}_{0}$ are Grassmann parameters.
Note that $\tilde{\bm{\delta}}_{\rm F}$ and $\tilde{\bm{\delta}}^{\dagger}_{\rm F}$
in (\ref{delta-F}) and (\ref{delta-Fdagger})
are related to $\tilde{\delta}_{\rm F}$ and $\tilde{\delta}_{\rm F}^{\dagger}$ as
$\tilde{\delta}_{\rm F} = \zeta_{0} \tilde{\bm{\delta}}_{\rm F}$,
$\tilde{\delta}^{\dagger}_{\rm F} 
= \zeta^{\dagger}_{0} \tilde{\bm{\delta}}^{\dagger}_{\rm F}$.

The system contains negative norm states originated from
$c_{\varphi}$, $c_{\varphi}^{\dagger}$ and $C_i^{\pm}$.
In the presence of negative norm states,
the probability interpretation cannot be endured.
To formulate our model in a consistent manner,
we use a feature that {\it conserved charges can be, in general, set to be zero
as subsidiary conditions}.
We impose the following subsidiary conditions on states by hand,
\begin{eqnarray}
N_{A} |{\rm phys}\rangle = 0,~~ N_{B} |{\rm phys}\rangle = 0,~~
Q_{\rm F} |{\rm phys}\rangle = 0,~~
Q_{\rm F}^{\dagger} |{\rm phys}\rangle = 0.
\label{Phys}
\end{eqnarray}
In appendix A, we point out that subsidiary conditions
corresponding to (\ref{Phys}) can be realized as remnants of local symmetries
in a specific case. 

\subsection{Unitarity}

Let us study the unitarity of physical $S$ matrix in our system,
using the Lagrangian density of free fields,
\begin{eqnarray}
\mathcal{L}_{0} 
= \partial_{\mu} \varphi^{\dagger} \partial^{\mu} \varphi - m^2 \varphi^{\dagger} \varphi
+ \partial_{\mu} c_{\varphi}^{\dagger} \partial^{\mu} c_{\varphi} 
- m^2 c_{\varphi}^{\dagger} c_{\varphi} 
- 2 \partial_{\mu} A_i \partial^{\mu} B^{i}
- \partial_{\mu} C_i^{+} \partial^{\mu} C^{-i},
\label{L-0}
\end{eqnarray}
where the gauge fixing conditions (\ref{gf}) are imposed on.
The $\mathcal{L}_{0}$ describes the behavior of asymptotic fields
of Heisenberg operators in  $\mathcal{L} = \mathcal{L}_{\rm M} + \mathcal{L}_{\rm G}$.

From (\ref{L-0}), free field equations for 
$\varphi$, $\varphi^{\dagger}$, $c_{\varphi}$, $c_{\varphi}^{\dagger}$,
$A_i$, $B_i$ and $C_i^{\pm}$ are derived.
By solving the Klein-Gordon equations, 
we obtain the solutions
\begin{eqnarray}
&~& \varphi(x) = \int \frac{d^3k}{\sqrt{(2\pi)^3 2k_0}}
\left(a(\bm{k}) e^{-i k x} + b^{\dagger}(\bm{k}) e^{i k x}\right),
\label{varphi-sol}\\
&~& \varphi^{\dagger}(x) = \int \frac{d^3k}{\sqrt{(2\pi)^3 2k_0}}
\left(a^{\dagger}(\bm{k}) e^{i k x} + b (\bm{k}) e^{-i k x}\right),
\label{varphi-dagger-sol}\\
&~& \pi(x) = i \int d^3k \sqrt{\frac{k_0}{2 (2\pi)^3}}
\left(a^{\dagger}(\bm{k}) e^{i k x} - b (\bm{k}) e^{-i k x}\right),
\label{pi-sol}\\
&~& \pi^{\dagger}(x) = - i \int d^3k \sqrt{\frac{k_0}{2 (2\pi)^3}}
\left(a(\bm{k}) e^{-i k x} - b^{\dagger} (\bm{k}) e^{i k x}\right),
\label{pi-dagger-sol}\\
&~& c_{\varphi}(x) = \int \frac{d^3k}{\sqrt{(2\pi)^3 2k_0}}
\left(c(\bm{k}) e^{-i k x} + d^{\dagger}(\bm{k}) e^{i k x}\right),
\label{c-sol}\\
&~& c_{\varphi}^{\dagger}(x) = \int \frac{d^3k}{\sqrt{(2\pi)^3 2k_0}}
\left(c^{\dagger}(\bm{k}) e^{i k x} + d (\bm{k}) e^{-i k x}\right),
\label{c-dagger-sol}\\
&~& \pi_{c_{\varphi}}(x) = i \int d^3k \sqrt{\frac{k_0}{2 (2\pi)^3}}
\left(c^{\dagger}(\bm{k}) e^{i k x} - d (\bm{k}) e^{-i k x}\right),
\label{pi-c-sol}\\
&~& \pi_{c_{\varphi}}^{\dagger}(x) = - i \int d^3k \sqrt{\frac{k_0}{2 (2\pi)^3}}
\left(c(\bm{k}) e^{-i k x} - d^{\dagger} (\bm{k}) e^{i k x}\right),
\label{pi-c-dagger-sol}
\end{eqnarray}
where $k_0 = \sqrt{\bm{k}^2 + m^2}$ and $kx = k^{\mu} x_{\mu}$.

In the same way, by solving the free Maxwell equations,
we obtain the solutions,
\begin{eqnarray}
&~& A_{i}(x) = \int \frac{d^3k}{\sqrt{(2\pi)^3 2k_0}}
\left(\varepsilon_i^{\alpha} a_{\alpha}(\bm{k}) e^{-i k x} 
+ \varepsilon_i^{*\alpha} a_{\alpha}^{\dagger}(\bm{k}) e^{i k x}\right),
\label{Ai-sol}\\
&~& B_{i}(x) = \int \frac{d^3k}{\sqrt{(2\pi)^3 2k_0}}
\left(\varepsilon_i^{\alpha} b_{\alpha}(\bm{k}) e^{-i k x} 
+ \varepsilon_i^{*\alpha} b_{\alpha}^{\dagger}(\bm{k}) e^{i k x}\right),
\label{Bi-sol}\\
&~& C_{i}^{+}(x) = \int \frac{d^3k}{\sqrt{(2\pi)^3 2k_0}}
\left(\varepsilon_i^{\alpha} c_{\alpha}(\bm{k}) e^{-i k x} 
+ \varepsilon_i^{*\alpha} d_{\alpha}^{\dagger}(\bm{k}) e^{i k x}\right),
\label{Ci+-sol}\\
&~& C_{i}^{-}(x) = \int \frac{d^3k}{\sqrt{(2\pi)^3 2k_0}}
\left(\varepsilon_i^{*\alpha} c_{\alpha}^{\dagger}(\bm{k}) e^{i k x}
+ \varepsilon_i^{\alpha} d_{\alpha}(\bm{k}) e^{-i k x}\right),
\label{Ci--sol}\\
&~& \Pi_{A}^{i}(x) = 2 i \int d^3k \sqrt{\frac{k_0}{2 (2\pi)^3}}
\left(\varepsilon_i^{*\alpha} b_{\alpha}^{\dagger}(\bm{k}) e^{i k x}
- \varepsilon_i^{\alpha} b_{\alpha}(\bm{k}) e^{-i k x}\right),
\label{PiAi-sol}\\
&~& \Pi_{B}^{i}(x) = 2 i \int d^3k \sqrt{\frac{k_0}{2 (2\pi)^3}}
\left(\varepsilon_i^{*\alpha} a_{\alpha}^{\dagger}(\bm{k}) e^{i k x}
- \varepsilon_i^{\alpha} a_{\alpha}(\bm{k}) e^{-i k x}\right),
\label{PiBi-sol}\\
&~& \Pi_{C}^{+i}(x) = i \int d^3k \sqrt{\frac{k_0}{2 (2\pi)^3}}
\left(\varepsilon_i^{*\alpha} c_{\alpha}^{\dagger}(\bm{k}) e^{i k x} 
- \varepsilon_i^{\alpha} d_{\alpha}(\bm{k}) e^{-i k x}\right),
\label{Pii+-sol}\\
&~& \Pi_{C}^{-i}(x) = - i \int d^3k \sqrt{\frac{k_0}{2 (2\pi)^3}}
\left(\varepsilon_i^{\alpha} c_{\alpha}(\bm{k}) e^{-i k x}
- \varepsilon_i^{*\alpha} d_{\alpha}^{\dagger}(\bm{k}) e^{i k x}\right),
\label{Pii--sol}
\end{eqnarray}
where $k_0 = |\bm{k}|$ and $\varepsilon_i^{\alpha}$ are polarization vectors
satisfying the relations,
\begin{eqnarray}
k_i \varepsilon_i^{\alpha} = 0,~~
\varepsilon_i^{\alpha} \varepsilon^{*i\alpha'} = \delta_{\alpha\alpha'},~~
\sum_{\alpha} \varepsilon_i^{\alpha} \varepsilon^{*j\alpha}
= \delta_i^j - \frac{k_i k^j}{\bm{k}^2}~.
\label{polarization}
\end{eqnarray}
The index $\alpha$ represents the helicity of gauge fields.

By imposing the same type of relations
as (\ref{CCR-varphi}) -- (\ref{CCR-C-}), 
we have the relations,
\begin{eqnarray}
&~& [a(\bm{k}), a^{\dagger}(\bm{l})] = \delta^3(\bm{k}-\bm{l}),~~ 
[b(\bm{k}), b^{\dagger}(\bm{l})] = \delta^3(\bm{k}-\bm{l}), 
\label{CCR-ab-varphi}\\
&~& \{c(\bm{k}), c^{\dagger}(\bm{l})\} = \delta^3(\bm{k}-\bm{l}),~~
\{d(\bm{k}), d^{\dagger}(\bm{l})\} = - \delta^3(\bm{k}-\bm{l}), 
\label{CCR-cd-c}\\
&~& [a_{\alpha}(\bm{k}), b_{\alpha'}^{\dagger}(\bm{l})] 
= - \frac{1}{2}\delta_{\alpha\alpha'} \delta^3(\bm{k}-\bm{l}),~~
[b_{\alpha}(\bm{k}), a_{\alpha'}^{\dagger}(\bm{l})] 
= - \frac{1}{2}\delta_{\alpha\alpha'} \delta^3(\bm{k}-\bm{l}),~~
\label{CCR-ab-AB}\\
&~& \{c_{\alpha}(\bm{k}),  c_{\alpha'}^{\dagger}(\bm{l})\} 
= \delta_{\alpha\alpha'} \delta^3(\bm{k}-\bm{l}),~~
\{d_{\alpha}(\bm{k}), d_{\alpha}^{\dagger}(\bm{l})\} 
= - \delta_{\alpha\alpha'} \delta^3(\bm{k}-\bm{l}), 
\label{CCR-cd-C}
\end{eqnarray}
and others are zero.

The states in the Fock space are constructed by acting
the creation operators $a^{\dagger}(\bm{k})$, $b^{\dagger}(\bm{k})$,
$c^{\dagger}(\bm{k})$, $d^{\dagger}(\bm{k})$,
$a_{\alpha}^{\dagger}(\bm{k})$, $b_{\alpha}^{\dagger}(\bm{k})$, 
$c_{\alpha}^{\dagger}(\bm{k})$ and $d_{\alpha}^{\dagger}(\bm{k})$
on the vacuum state $| 0 \rangle$,
where $| 0 \rangle$ is defined by the conditions $a(\bm{k})| 0 \rangle = 0$,
$b(\bm{k})| 0 \rangle = 0$, $c(\bm{k})| 0 \rangle = 0$, $d(\bm{k})| 0 \rangle = 0$,
$a_{\alpha}(\bm{k}) | 0 \rangle = 0$, $b_{\alpha}(\bm{k}) | 0 \rangle = 0$,
$c_{\alpha}(\bm{k}) | 0 \rangle = 0$ and $d_{\alpha}(\bm{k}) | 0 \rangle = 0$.

We impose the following subsidiary conditions on states to select physical states,
\begin{eqnarray}
N_{A} |{\rm phys}\rangle = 0,~~ N_{B} |{\rm phys}\rangle = 0,~~
Q_{\rm F} |{\rm phys}\rangle = 0,~~
Q_{\rm F}^{\dagger} |{\rm phys}\rangle = 0.
\label{Phys2}
\end{eqnarray}
Note that $Q_{\rm F}^{\dagger} |{\rm phys}\rangle = 0$ means $\langle {\rm phys}|Q_{\rm F}=0$.
We find that all states, 
except for the vacuum state, are
unphysical because they do not satisfy (\ref{Phys2}).
This feature is understood as a counterpart of the quartet mechanism~\cite{K&O1,K&O2}.
The projection operator $P^{(n)}$ on the states with $n$ particles
is given by
\begin{eqnarray}
&~& P^{(n)} = \frac{1}{n} \left\{a^{\dagger} P^{(n-1)} a + b^{\dagger} P^{(n-1)} b 
+ c^{\dagger} P^{(n-1)} c - d^{\dagger} P^{(n-1)} d \right.
\nonumber \\
&~& ~~~~~~~~~~~~ \left. 
+ \sum_{\alpha}\left(-2 a_{\alpha}^{\dagger} P^{(n-1)} b_{\alpha} 
- 2 b_{\alpha}^{\dagger} P^{(n-1)} a_{\alpha} 
+ c_{\alpha}^{\dagger} P^{(n-1)} c_{\alpha} 
- d_{\alpha}^{\dagger} P^{(n-1)} d_{\alpha} \right)\right\},
\label{P(n)}
\end{eqnarray}
where $n \ge 1$ and
we omit $\bm{k}$, for simplicity.
Using the transformation properties,
\begin{eqnarray}
&~& \tilde{\bm{\delta}}_{\rm F} a = - c,~~
\tilde{\bm{\delta}}_{\rm F} a^{\dagger} = 0,~~ 
\tilde{\bm{\delta}}_{\rm F} b = 0,~~
\tilde{\bm{\delta}}_{\rm F} b^{\dagger} = - d^{\dagger},~~
\nonumber \\
&~& \tilde{\bm{\delta}}_{\rm F} c  = 0,~~ 
\tilde{\bm{\delta}}_{\rm F} c^{\dagger} = a^{\dagger},~~
\tilde{\bm{\delta}}_{\rm F} d = b,~~
\tilde{\bm{\delta}}_{\rm F} d^{\dagger} = 0,
\nonumber \\
&~& \tilde{\bm{\delta}}_{\rm F} a_{\alpha} = - i d_{\alpha},~~
\tilde{\bm{\delta}}_{\rm F} a_{\alpha}^{\dagger} = - i c_{\alpha}^{\dagger},~~ 
\tilde{\bm{\delta}}_{\rm F} b_{\alpha} = 0,~~
\tilde{\bm{\delta}}_{\rm F} b_{\alpha}^{\dagger} = 0,~~
\nonumber \\
&~& \tilde{\bm{\delta}}_{\rm F} c_{\alpha}  = 2 i b_{\alpha},~~ 
\tilde{\bm{\delta}}_{\rm F} c_{\alpha}^{\dagger} = 0,~~
\tilde{\bm{\delta}}_{\rm F} d_{\alpha} = 0,~~
\tilde{\bm{\delta}}_{\rm F} d_{\alpha}^{\dagger} = 2 i b_{\alpha}^{\dagger},
\label{delta-F-abcd}
\end{eqnarray}
$P^{(n)}$ is written in a simple form as
\begin{eqnarray}
P^{(n)} =  i \left\{Q_{\rm F}, R^{(n)}\right\}~,
\label{P(n)2}
\end{eqnarray}
where $R^{(n)}$ is given by
\begin{eqnarray}
R^{(n)} = \frac{1}{n} \left\{
c^{\dagger} P^{(n-1)} a + b^{\dagger} P^{(n-1)} d
+ i \sum_{\alpha} \left(a_{\alpha}^{\dagger} P^{(n-1)} c_{\alpha} 
+ d_{\alpha}^{\dagger} P^{(n-1)} a_{\alpha}\right)\right\}.
\label{R(n)}
\end{eqnarray}
From (\ref{P(n)2}), we find that any state with $n \ge 1$ is unphysical from
$\langle {\rm phys}|P^{(n)}|{\rm phys}\rangle = 0$.
Then, we understand that every field becomes unphysical,
and only $|0 \rangle$ remains as the physical state.
This is also regarded as a field theoretical version 
of the Parisi-Sourlas mechanism~\cite{P&S}.

The system is also formulated using hermitian fermionic charges
defined by $Q_1 \equiv Q_{\rm F} + Q_{\rm F}^{\dagger}$ and 
$Q_2 \equiv i(Q_{\rm F} - Q_{\rm F}^{\dagger})$.
They satisfy the relations $Q_1 Q_2 + Q_2 Q_1 = 0$,
${Q_1}^2 = N_{A}$ and ${Q_2}^2 = N_{A}$.
Though $Q_1$, $Q_2$ and $N_{A}$
form elements of the $N=2$ (quantum mechanical) 
SUSY algebra~\cite{Witten},
our system does not possess the space-time SUSY
because $N_{A}$ is not our Hamiltonian
but the $U(1)$ charge $N_{A}$.
Only the vacuum state is selected as the physical states
by imposing the following subsidiary conditions on states, 
in place of (\ref{Phys2}),
\begin{eqnarray}
N_{A} |{\rm phys}\rangle = 0,~~ N_{B} |{\rm phys}\rangle = 0,~~
Q_{1} |{\rm phys}\rangle = 0,~~
Q_{2} |{\rm phys}\rangle = 0.
\label{Phys-Q12}
\end{eqnarray}
It is also understood that our fermionic symmetries are different from 
the space-time SUSY,
from the fact that $Q_1$ and $Q_2$ are scalar charges.
They are also different from the BRST symmetry,
as seen from the algebraic relations among charges.

The system with spinor and gauge fields described by 
$\mathcal{L}^{\rm sp} = \mathcal{L}_{\rm M}^{\rm sp} + \mathcal{L}_{\rm G}$
is also quantized, in a similar way.
We find that the theory becomes harmless but empty leaving the vacuum state alone
as the physical state, after imposing subsidiary conditions 
corresponding to (\ref{Phys}).

\subsection{BRST symmetry}

Our system has local symmetries,
and it is quantized by the Faddeev-Popov (FP) method.
In order to add the gauge fixing conditions to the Lagrangian,
several fields corresponding to FP ghost and anti-ghost fields 
and auxiliary fields called Nakanishi-Lautrup (NL) fields are introduced.
Then, the system is described on the extended phase space
and has a global symmetry called the BRST symmetry. 
We present the gauge-fixed Lagrangian density
and study the BRST transformation properties.

According to the usual procedure,
the Lagrangian density containing the gauge fixing terms
and FP ghost terms is constructed as
\begin{eqnarray}
&~& \mathcal{L}_{\rm T}= \mathcal{L}_{\rm M} + \mathcal{L}_{\rm G}
+ \mathcal{L}_{\rm gf} + \mathcal{L}_{\rm FP},
\nonumber\\
&~& \mathcal{L}_{\rm gf}
= - \partial_{\mu}b_A~A^{\mu} -  \partial_{\mu}b_B~B^{\mu} 
+ C^{+\mu} \partial_{\mu}\phi_c
+ \partial_{\mu}\phi_c^{\dagger}~C^{-\mu}
\nonumber \\
&~& ~~~~~~~~~~~ + \frac{1}{2}\alpha (b_A^2 + b_B^2 + 2 \phi_c^{\dagger} \phi_c),
\label{L-gf}\\
&~& \mathcal{L}_{\rm FP}
=- i \partial_{\mu} \overline{c}_A (\partial^{\mu} c_A
 - i g \phi C^{-\mu} + i g \phi^{\dagger} C^{+\mu})
 - i \partial_{\mu} \overline{c}_B~\partial^{\mu} c_B
\nonumber \\
&~& ~~~~~~~~~~~~ + i (\partial^{\mu} \phi
 - 2 i g c_B C^{+\mu} + 2 i g \phi B^{\mu}) \partial_{\mu} \overline{\phi}
\nonumber \\
&~& ~~~~~~~~~~~~ - i \partial_{\mu} \overline{\phi}^{\dagger} (\partial^{\mu} \phi^{\dagger}
 - 2 i g c_B C^{-\mu} - 2 i g \phi^{\dagger} B^{\mu}),
\label{L-FP}
\end{eqnarray}
where $c_A$, $c_B$, $\phi$ and $\phi^{\dagger}$ are FP ghosts,
$\overline{c}_A$, $\overline{c}_B$, $\overline{\phi}$ and $\overline{\phi}^{\dagger}$
are FP anti-ghosts,
$b_A$, $b_B$, $\phi_c$ and $\phi_c^{\dagger}$
are NL fields,
and $\alpha$ is a gauge parameter.
These fields are scalar fields.
$c_A$, $c_B$, $\overline{c}_A$ and $\overline{c}_B$ are fermionic,
and $b_A$ and $b_B$ are bosonic.
In contrast, $\phi$, $\phi^{\dagger}$, $\overline{\phi}$ and $\overline{\phi}^{\dagger}$
are bosonic, 
and $\phi_c$ and $\phi_c^{\dagger}$ are fermionic
because the relevant symmetries are fermionic.

The $\mathcal{L}_{\rm T}$ is invariant 
under the BRST transformation,
\begin{eqnarray}
&~& \bm{\delta}_{\mbox{\tiny BRST}} \varphi 
= - i g c_A \varphi - i g c_B \varphi  - g \phi c_{\varphi},~~
 \bm{\delta}_{\mbox{\tiny BRST}} \varphi^{\dagger} 
= i g c_A \varphi^{\dagger} + i g c_B \varphi^{\dagger}  - g \phi^{\dagger} c_{\varphi}^{\dagger},~~
\nonumber \\
&~& \bm{\delta}_{\mbox{\tiny BRST}} c_{\varphi} 
= - i g c_A c_{\varphi} + i g c_B c_{\varphi} - g \phi^{\dagger} \varphi,~~
\bm{\delta}_{\mbox{\tiny BRST}} c_{\varphi}^{\dagger} 
= i g c_A c_{\varphi}^{\dagger} -  i g c_B c_{\varphi}^{\dagger} + g \phi \varphi^{\dagger},
\nonumber \\
&~& \bm{\delta}_{\mbox{\tiny BRST}}  c_A = -i g \phi^{\dagger} \phi,~~
\bm{\delta}_{\mbox{\tiny BRST}}  c_B = 0,~~
\bm{\delta}_{\mbox{\tiny BRST}} \phi = - 2 i g c_B \phi,~~
\bm{\delta}_{\mbox{\tiny BRST}}  \phi^{\dagger} = 2 i g c_B \phi^{\dagger},~~
\nonumber \\
&~& \bm{\delta}_{\mbox{\tiny BRST}} A_{\mu} 
= \partial_{\mu} c_A - i g \phi C_{\mu}^{-} + i g \phi^{\dagger} C_{\mu}^{+},~~
\bm{\delta}_{\mbox{\tiny BRST}} B_{\mu} = \partial_{\mu} c_B,~~
\nonumber \\
&~& \bm{\delta}_{\mbox{\tiny BRST}} C_{\mu}^{+} 
= - 2 i g c_B C_{\mu}^+ + 2 i g \phi B_{\mu} + \partial_{\mu} \phi,~~
\bm{\delta}_{\rm B}  C_{\mu}^{-} 
= 2 i g c_B C_{\mu}^- + 2 i g \phi^{\dagger} B_{\mu} - \partial_{\mu} \phi^{\dagger},
\nonumber \\
&~& \bm{\delta}_{\mbox{\tiny BRST}} \overline{c}_A = i b_A,~~
\bm{\delta}_{\mbox{\tiny BRST}} \overline{c}_B = i b_B,~~
\bm{\delta}_{\mbox{\tiny BRST}} \overline{\phi} = i \phi_c,~~
\bm{\delta}_{\mbox{\tiny BRST}} \overline{\phi}^{\dagger} = - i {\phi}_c^{\dagger},~~
\nonumber \\
&~& \bm{\delta}_{\mbox{\tiny BRST}} b_A = 0,~~
\bm{\delta}_{\mbox{\tiny BRST}} b_B = 0,~~
\bm{\delta}_{\mbox{\tiny BRST}} \phi_c = 0,~~
\bm{\delta}_{\mbox{\tiny BRST}} \phi_c^{\dagger} = 0,
\label{delta-BRST}
\end{eqnarray}
where the transformations for $\varphi$, $\varphi^{\dagger}$, $c_{\varphi}$,
$c_{\varphi}^{\dagger}$, $A_{\mu}$, $B_{\mu}$, $C_{\mu}^+$ and $C_{\mu}^-$
are obtained by regarding the sum of transformations
$\delta_{A} + \delta_{B} + \delta_{\rm F} + \delta_{\rm F}^{\dagger}$
as $\bm{\delta}_{\mbox{\tiny BRST}}$ and
replacing $\epsilon$, $\xi$, $\zeta$ and $\zeta^{\dagger}$
with $g c_A$, $g c_B$, $g \phi$ and $-g \phi^{\dagger}$,
and those for $c_A$, $c_B$, $\phi$ and $\phi^{\dagger}$
are determined by the requirement that 
$\bm{\delta}_{\mbox{\tiny BRST}}$ has a nilpotency property, i.e.,
${\bm{\delta}_{\mbox{\tiny BRST}}}^2 \mathcal{O} = 0$.

The sum of the gauge fixing terms
and FP ghost terms is simply written as
\begin{eqnarray}
&~& \mathcal{L}_{\rm gf} + \mathcal{L}_{\rm FP}
= i \bm{\delta}_{\mbox{\tiny BRST}} \bigl\{\partial_{\mu}\overline{c}_A~A^{\mu} 
+ \partial_{\mu}\overline{c}_B~B^{\mu} 
+ C^{+\mu} \partial_{\mu}\overline{\phi}
+ \partial_{\mu}\overline{\phi}^{\dagger}~C^{-\mu}
\nonumber\\
&~& ~~~~~~~~~~~~~~~~~~~~~~~~~~~~~~~~~~~~ 
- \frac{1}{2}\alpha (\overline{c}_A b_A + \overline{c}_B b_B 
- \phi_c^{\dagger} \overline{\phi}  - \overline{\phi}^{\dagger} \phi_c)\bigr\}.
\label{L-fgFP}
\end{eqnarray}

According to the Noether procedure, 
the BRST current $J_{\mbox{\tiny BRST}}^{\mu}$ and
the BRST charge $Q_{\mbox{\tiny BRST}}$ are obtained as
\begin{eqnarray}
&~& J_{\mbox{\tiny BRST}}^{\mu} 
= b_A (\partial^{\mu} c_{A} - i g \phi C^{-\mu} + i g \phi^{\dagger} C^{+\mu})
- c_{A} \partial^{\mu} b_A
+ b_B \partial^{\mu} c_{B}~ - c_{B} \partial^{\mu} b_B
\nonumber \\
&~& ~~~~~~~~~~~~ 
- \phi_c (\partial^{\mu} \phi - 2 i g c_B C^{+\mu} + 2 i g \phi B^{\mu})
+ \phi \partial^{\mu} \phi_c
\nonumber \\
&~& ~~~~~~~~~~~~
- \phi_c^{\dagger} (\partial^{\mu} \phi^{\dagger} - 2 i g c_B C^{-\mu}
- 2 i g \phi^{\dagger} B^{\mu})
+ \phi^{\dagger} \partial^{\mu} \phi_c^{\dagger}
\nonumber \\
&~& ~~~~~~~~~~~~ 
- 2 g c_B \phi \partial^{\mu} \overline{\phi}
- 2 g c_B \phi^{\dagger} \partial^{\mu} \overline{\phi}^{\dagger}
- g \phi^{\dagger} \phi \partial^{\mu} \overline{c}_A
\nonumber \\
&~& ~~~~~~~~~~~~ 
- 2 \partial_{\nu} (c_A B^{\mu\nu}) - 2 \partial_{\nu} (c_B A^{\mu\nu})
- \partial_{\nu} (\phi C^{-\mu\nu}) - \partial_{\nu} (\phi^{\dagger} C^{+\mu\nu})
\label{J-BRST}
\end{eqnarray}
and
\begin{eqnarray}
&~& Q_{\mbox{\tiny BRST}}
\equiv \int d^3x J_{\mbox{\tiny BRST}}^{0}
= \int d^3x \bigl\{b_A  (\partial^{0} c_{A} - i g \phi C^{-0} + i g \phi^{\dagger} C^{+0})
 - c_{A} \partial^{0} b_A
\nonumber \\
&~& ~~~~~~~~~~~~
+ b_B \partial^{0} c_{B}~ - \partial^{0} b_B~c_{B} 
- \phi_c (\partial^{0} \phi - 2 i g c_B C^{+0} + 2 i g \phi B^{0})
+ \phi \partial^{0} \phi_c
\nonumber \\
&~& ~~~~~~~~~~~~
- \phi_c^{\dagger} (\partial^{0} \phi^{\dagger} - 2 i g c_B C^{-0}
- 2 i g \phi^{\dagger} B^{0})
+ \phi^{\dagger} \partial^{0} \phi_c^{\dagger}
\nonumber\\
&~& ~~~~~~~~~~~~ 
- 2 g c_B \phi \partial^{0} \overline{\phi}
- 2 g c_B \phi^{\dagger} \partial^{0} \overline{\phi}^{\dagger}
- g \phi^{\dagger} \phi \partial^{0} \overline{c}_A\bigr\},
\label{Q-BRST}
\end{eqnarray}
respectively. 
Here we use the field equations.
The BRST charge is a conserved charge ($d Q_{\mbox{\tiny BRST}}/dt = 0$),
and it has the nilpotency property such as ${Q_{\mbox{\tiny BRST}}}^2 = 0$.

By imposing the following subsidiary condition on states, 
\begin{eqnarray}
Q_{\mbox{\tiny BRST}} |{\rm phys}\rangle = 0,
\label{Phys-BRST}
\end{eqnarray}
it is shown that any negative states originated from
time and longitudinal components of
gauge fields as well as FP ghost and anti-ghost fields and NL fields
do not appear on the physical subspace, through the quartet mechanism.
There still exist negative norm states come from
$c_{\varphi}$, $c_{\varphi}^{\dagger}$ and $C_{\mu}^{\pm}$,
and it is necessary to impose additional conditions corresponding to
(\ref{Phys}) on states in order to project out such harmful states.

\section{Conclusions and discussions}

We have studied the quantization of systems with local particle-ghost symmetries.
The systems contain ordinary particles including gauge bosons
and their counterparts obeying different statistics.
There exist negative norm states come from fermionic scalar fields 
(or bosonic spinor fields)
and transverse components of fermionic gauge fields,
even after reducing the phase space due to 
the first class constraints and the gauge fixing conditions
or imposing the subsidiary condition concerning the BRST charge on states.
By imposing additional subsidiary conditions on states,
such negative norm states are projected out on the physical subspace
and the unitarity of systems hold.
The additional conditions can be originated from constraints 
in case that gauge fields have no dynamical degrees of freedom.

The systems considered are unrealistic if this goes on,
because they are empty leaving the vacuum state alone
as the physical state.
Then, one might think that it is better not to get deeply involved them.
Although they are still up in the air at present, but
there is a possibility that a formalism or concept itself is basically correct
and is useful to explain phenomena of elementary particles at a more fundamental level.
It is necessary to fully understand features of our particle-ghost symmetries,
in order to  appropriately apply them on a more microscopic system.

We make conjectures on some applications.
We suppose that particle-ghost symmetries exist 
and the system contains only a few states including the vacuum one
as physical states at an ultimate level.
Most physical particles might be released from
unphysical doublets that consist of particles and their ghost partners.
A release mechanism has been proposed 
based on the dimensional reduction by orbifolding~\cite{YK3}.

After the appearance of physical fields,
$Q_{\rm F}$-singlets and $Q_{\rm F}$-doublets coexist with exact fermionic symmetries.
The Lagrangian density is, in general, written in the form as
$\mathcal{L}_{\rm Total} = \mathcal{L}_{\rm S} + \mathcal{L}_{\rm D} + \mathcal{L}_{\rm mix}
= \mathcal{L}_{\rm S} 
+ \tilde{\bm{\delta}}_{\rm F} \tilde{\bm{\delta}}_{\rm F}^{\dagger} (\Delta \mathcal{L})$.
Here, $\mathcal{L}_{\rm S}$, $\mathcal{L}_{\rm D}$ 
and $\mathcal{L}_{\rm mix}$
stand for the Lagrangian density for $Q_{\rm F}$-singlets, 
$Q_{\rm F}$-doublets
and interactions between $Q_{\rm F}$-singlets and $Q_{\rm F}$-doublets.
Under the subsidiary conditions 
$N_{A} |{\rm phys}\rangle =0$,  $N_{B} |{\rm phys}\rangle =0$,
$Q_{\rm F} |{\rm phys}\rangle =0$ and
$Q_{\rm F}^{\dagger} |{\rm phys}\rangle =0$
on states,
all $Q_{\rm F}$-doublets become unphysical
and would not give any physical effects on $Q_{\rm F}$ singlets.
Because $Q_{\rm F}$ singlets would not receive any radiative corrections 
from $Q_{\rm F}$ doublets, the theory is free from the gauge hierarchy problem
if all heavy fields form $Q_{\rm F}$ doublets~\cite{YK1}.

The system seems to be same as that described by 
$\mathcal{L}_{\rm S}$ alone, and to be impossible to show 
the existence of $Q_{\rm F}$-doublets.
However, in a very special case, an indirect proof would be possible
through fingerprints left by symmetries in a fundamental theory.
The fingerprints are specific relations among parameters
such as a unification of coupling constants,
reflecting on underlying symmetries~\cite{YK1,YK5}. 

In most cases, our ghost fields require non-local interactions~\cite{YK1}
and the change of degrees of freedom
can occur in systems with infinite numbers of fields~\cite{YK3}.
Then, they might suggest that fundamental objects are not point particles
but extended objects such as strings and membranes.
Hence, it would be interesting to explore systems with particle-ghost symmetries
and their applications
in the framework of string theories.\footnote{
Objects called ghost D-branes have been introduced
as an extension of D-brane and their properties have been studied~\cite{OT,Tera}.
}

\section*{Acknowledgments}
This work was supported in part by scientific grants from the Ministry of Education, Culture,
Sports, Science and Technology under Grant No.~22540272.

\appendix

\section{System with auxiliary gauge fields}

Let us study the system without $\mathcal{L}_{\rm G}$ described by
\begin{eqnarray}
&~& \mathcal{L}_{\rm M}
= \bigl\{(\partial_{\mu} - i g A_{\mu} - i g B_{\mu}) \varphi^{\dagger} 
- g C_{\mu}^{-} c_{\varphi}^{\dagger}\bigr\}
\bigl\{(\partial^{\mu} + i g A^{\mu} + i g B^{\mu}) \varphi
+ g C^{+\mu} c_{\varphi}\bigr\} 
\nonumber\\
&~& ~~~~~~~~~~~ 
+ \bigl\{(\partial_{\mu} - ig A_{\mu} + i g B_{\mu}) c_{\varphi}^{\dagger} 
- g C_{\mu}^{+} \varphi^{\dagger}\bigr\}
\bigl\{(\partial^{\mu} + i g A^{\mu} - i g B^{\mu}) c_{\varphi}
- g C^{-\mu} \varphi\bigr\} 
\nonumber\\
&~& ~~~~~~~~~~~ 
 - m^2 \varphi^{\dagger} \varphi - m^2 c_{\varphi}^{\dagger} c_{\varphi}.
\label{L-M-again}
\end{eqnarray}
In this case, gauge fields do not have any dynamical degrees of freedom,
and are regarded as auxiliary fields.
The conjugate momenta of $\varphi$, $\varphi^{\dagger}$,
$c_{\varphi}$ and $c_{\varphi}^{\dagger}$ are same as those 
obtained in (\ref{pi}) -- (\ref{pi-c-dagger}).
The conjugate momenta of $A_{\mu}$, $B_{\mu}$, $C_{\mu}^+$
and $C_{\mu}^-$ become constraints,
\begin{eqnarray}
\Pi_{A}^{\mu} = 0,~~ \Pi_{B}^{\mu} = 0,~~
\Pi_{C}^{+\mu} = 0,~~
\Pi_{C}^{-\mu} = 0.
\label{Pi-ABC}
\end{eqnarray}

Using the Legendre transformation,
the Hamiltonian density is obtained as
\begin{eqnarray}
&~& \mathcal{H}_{\rm M} = \pi \dot{\varphi} 
+ \dot{\varphi}^{\dagger}\pi^{\dagger} + \pi_{c_{\varphi}} \dot{c}_{\varphi} 
+ \dot{c}_{\varphi}^{\dagger} \pi_{c_{\varphi}}^{\dagger}
+ \dot{A}_{\mu} \Pi_{A}^{\mu}  + \Pi_{B}^{\mu} \dot{B}_{\mu}
+ \dot{C}_{\mu}^{+} \Pi_{C}^{+\mu} + \Pi_{C}^{-\mu} \dot{C}_{\mu}^{-}
- \mathcal{L} 
\nonumber \\
&~& ~~~~~~~~~~~~ + \lambda_{A\mu} \Pi_{A}^{\mu} + \Pi_{B}^{\mu} \lambda_{B\mu} 
+ \lambda_{C\mu}^{+} \Pi_{C}^{+\mu} + \Pi_{C}^{-\mu} \lambda_{C\mu}^{-}
\nonumber \\
&~& ~~~~~~~~~ = \pi \pi^{\dagger} + \pi_{c_{\varphi}} \pi_{c_{\varphi}}^{\dagger} 
+ (D_i \Phi)^{\dagger} (D^i \Phi) + m^2 \Phi^{\dagger} \Phi
\nonumber \\
&~& ~~~~~~~~~~~~ - i g A_0 (\pi \varphi - \varphi^{\dagger} \pi^{\dagger} 
+ \pi_{c_{\varphi}} c_{\varphi} - c_{\varphi}^{\dagger} \pi_{c_{\varphi}}^{\dagger})
- i g B_0 (\pi \varphi - \varphi^{\dagger} \pi^{\dagger} 
- \pi_{c_{\varphi}} c_{\varphi} + c_{\varphi}^{\dagger} \pi_{c_{\varphi}}^{\dagger})
\nonumber \\
&~& ~~~~~~~~~~~~ - g C_0^{+} (\pi c_{\varphi} - \varphi^{\dagger} \pi_{c_{\varphi}}^{\dagger})
 - g (c_{\varphi}^{\dagger} \pi^{\dagger} - \pi_{c_{\varphi}} \varphi) C_0^{-}
\nonumber \\
&~& ~~~~~~~~~~~~ + \lambda_{A\mu} \Pi_{A}^{\mu} + \Pi_{B}^{\mu} \lambda_{B\mu} 
+ \lambda_{C\mu}^{+} \Pi_{C}^{+\mu} + \Pi_{C}^{-\mu} \lambda_{C\mu}^{-},
\label{H-non}
\end{eqnarray}
where $\lambda_{A\mu}$, $\lambda_{B\mu}$, 
$\lambda_{C\mu}^{+}$ and $\lambda_{C\mu}^{-}$
are Lagrange multipliers.

Secondary constraints are obtained as
\begin{eqnarray}
&~& \frac{d\Pi_{A}^{\mu}}{dt} = \left\{\Pi_{A}^{\mu}, H_{\rm M}\right\}_{\rm PB} 
= g j_A^{\mu} = 0,~~
\label{secondaryA-non}\\
&~& \frac{d \Pi_{B}^{\mu}}{dt} = \left\{\Pi_{B}^{\mu}, H_{\rm M}\right\}_{\rm PB} 
= g j_B^{\mu} = 0,~~
\label{secondaryB-non}\\
&~& \frac{d \Pi_{C}^{+\mu}}{dt} = \left\{\Pi_{C}^{+\mu}, H_{\rm M}\right\}_{\rm PB} 
= g j_{C}^{+\mu} = 0,~~
\label{secondaryC+-non}\\
&~& \frac{d \Pi_{C}^{-\mu}}{dt} = \left\{\Pi_{C}^{-\mu}, H_{\rm M}\right\}_{\rm PB} 
= g j_{C}^{-\mu} = 0,
\label{secondaryC--non}
\end{eqnarray}
where $H_{\rm M}$ is the Hamiltonian $H_{\rm M} = \int \mathcal{H}_{\rm M} d^3x$,
and $j_A^{\mu}$, $j_{B}^{\mu}$, $j_{C}^{+\mu}$ and $j_{C}^{-\mu}$
are the currents of $U(1)$ and fermionic symmetries given by
\begin{eqnarray}
\hspace{-1cm}&~& j_{A}^{0} = i \left(\pi \varphi - \varphi^{\dagger} \pi^{\dagger}
+ \pi_{c_\varphi} c_{\varphi} 
 - c_{\varphi}^{\dagger} \pi_{c_{\varphi}}^{\dagger}\right),~~
\nonumber \\
\hspace{-1cm}&~& j_{A}^{i} = i \bigl[ 
\bigl\{(\partial^{i} - i g A^{i} - i g B^{i}) \varphi^{\dagger} - g C^{-i} c_{\varphi}^{\dagger}\bigr\} \varphi
- \varphi^{\dagger}
\bigl\{(\partial^{i} + i g A^{i} + i g B^{i}) \varphi + g C^{+i} c_{\varphi}\bigr\} 
\nonumber \\
\hspace{-1cm}&~& ~~~~~~~~ + \bigl\{(\partial^{i} + i g A^{i} - i g B^{i}) c_{\varphi}
- g C^{-i} \varphi\bigr\} c_{\varphi}
- c_{\varphi}^{\dagger} 
\bigl\{(\partial^{i} + i g A^{i} - i g B^{i}) c_{\varphi} - g C^{-i} \varphi\bigr\}\bigr],
\label{jAmu}\\
\hspace{-1cm}&~& j_{B}^{0} = i \left(\pi \varphi - \varphi^{\dagger} \pi^{\dagger}
- \pi_{c_\varphi} c_{\varphi} 
+ c_{\varphi}^{\dagger} \pi_{c_{\varphi}}^{\dagger}\right),~~
\nonumber \\
\hspace{-1cm}&~& j_{B}^{i} = i \bigl[ 
\bigl\{(\partial^{i} - i g A^{i} - i g B^{i}) \varphi^{\dagger} - g C^{-i} c_{\varphi}^{\dagger}\bigr\} \varphi
- \varphi^{\dagger}
\bigl\{(\partial^{i} + i g A^{i} + i g B^{i}) \varphi + g C^{+i} c_{\varphi}\bigr\} 
\nonumber \\
\hspace{-1cm}&~& ~~~~~~~~ - \bigl\{(\partial^{i} + i g A^{i} - i g B^{i}) c_{\varphi}
- g C^{-i} \varphi\bigr\} c_{\varphi}
+ c_{\varphi}^{\dagger} 
\bigl\{(\partial^{i} + i g A^{i} - i g B^{i}) c_{\varphi} - g C^{-i} \varphi\bigr\}\bigr],
\label{jBmu}\\
\hspace{-1cm}&~& j_{C}^{+0} = \pi c_{\varphi} 
- \varphi^{\dagger} \pi_{c_{\varphi}}^{\dagger},~~
\nonumber \\
\hspace{-1cm}&~& j_{C}^{+i} =  \bigl\{(\partial^{i} - i g A^{i} - i g B^{i}) \varphi^{\dagger} 
- g C^{-i} c_{\varphi}^{\dagger}\bigr\} c_{\varphi}
- \varphi^{\dagger} 
\bigl\{(\partial^{i} + i g A^{i} - i g B^{i}) c_{\varphi}
- g C^{-i} \varphi\bigr\},
\label{jC+mu}\\
\hspace{-1cm}&~& j_{C}^{-0} = c_{\varphi}^{\dagger} \pi^{\dagger} 
- \pi_{c_{\varphi}} \varphi,~~
\nonumber \\
\hspace{-1cm}&~& j_{C}^{-i} = c_{\varphi}^{\dagger}
\bigl\{(\partial^{i} + i g A^{i} + i g B^{i}) \varphi + g C^{+i} c_{\varphi}\bigr\} 
- \bigl\{(\partial^{i} - i g A^{i} + i g B^{i}) c_{\varphi}^{\dagger} 
- g C^{+i} \varphi^{\dagger}\bigr\} \varphi.
\label{jC-mu}
\end{eqnarray}
In the same way, tertiary constraints are obtained as
\begin{eqnarray}
&~& \frac{d j_{A}^{0}}{dt} = \left\{j_{A}^{0}, H_{\rm M}\right\}_{\rm PB} 
= - \partial_i j_{A}^{i} = 0,
\label{tertiaryA}\\
&~& \frac{d j_{B}^{0}}{dt} = \left\{j_{B}^{0}, H_{\rm M}\right\}_{\rm PB} 
= - \partial_i j_{B}^{i} = 0,
\label{tertiaryB}\\
&~& \frac{d j_{C}^{+0}}{dt} = \left\{j_{C}^{+0}, H_{\rm M}\right\}_{\rm PB} 
= - \partial_i j_{C}^{+i} = 0,
\label{tertiaryC+}\\
&~& \frac{d j_{C}^{-0}}{dt} = \left\{j_{C}^{-0}, H_{\rm M}\right\}_{\rm PB} 
= - \partial_i j_{C}^{-i} = 0,
\label{tertiaryC-}
\end{eqnarray}
from the invariance under the time evolution of
$j_{A}^{0} = 0$, $j_{B}^{0} = 0$, $j_{C}^{+0} = 0$ and $j_{C}^{-0} = 0$.
On the other hand, the conditions 
$d j_{A}^{i}/dt = \left\{j_{A}^{i}, H_{\rm M}\right\}_{\rm PB} = 0$,
$d j_{B}^{i}/dt = \left\{j_{B}^{i}, H_{\rm M}\right\}_{\rm PB} = 0$,
$d j_{C}^{+i}/dt = \left\{j_{C}^{+i}, H_{\rm M}\right\}_{\rm PB} = 0$
and $d j_{C}^{-i}/dt = \left\{j_{C}^{-i}, H_{\rm M}\right\}_{\rm PB} = 0$
are not new constraints but the relations to determine $\lambda_{Ai}$, $\lambda_{Bi}$,
$\lambda_{Ci}^{+}$ and $\lambda_{Ci}^{-}$.
Furthermore, new constraints do not appear from the conditions
$d (\partial_i j_{A}^{i})/dt = 0$, $d (\partial_i j_{B}^{i})/dt = 0$,
$d (\partial_i j_{C}^{+i})/dt = 0$ and $d (\partial_i j_{C}^{-i})/dt = 0$.

The constraints are classified into the first class ones
\begin{eqnarray}
\Pi_{A}^{0} = 0,~~ \Pi_{B}^{0}  = 0,~~ \Pi_{C}^{+0} = 0,~~ \Pi_{C}^{-0} = 0
\label{first-non}
\end{eqnarray}
and the second class ones
\begin{eqnarray}
&~&\Pi_{A}^{i} = 0,~~ \Pi_{B}^{i}  = 0,~~ \Pi_{C}^{+i} = 0,~~ \Pi_{C}^{-i}  = 0,~~
\nonumber\\
&~& j_{A}^{i} = 0,~~ j_{B}^{i} = 0,~~ j_{C}^{+i} = 0,~~ j_{C}^{-i} = 0,~~ 
\nonumber\\
&~& j_{A}^0 = 0,~~  j_{B}^{0} = 0,~~ j_{C}^{+0} = 0,~~ j_{C}^{-0} = 0,~~
\nonumber\\
&~& \partial_i j_{A}^{i} = 0,~~ \partial_i j_{B}^{i} = 0,~~
\partial_i j_{C}^{+i} = 0,~~ \partial_i j_{C}^{-i} = 0.
\label{second-non}
\end{eqnarray}
The determinant
of Poisson bracket between second class ones $\{\phi_{\rm 2nd}^{a}\}$
does not vanish on constraints.

Using $j_{A}^0$, $j_{B}^{0}$, $j_{C}^{+0}$ and $j_{C}^{-0}$, 
the conserved $U(1)$ and fermionic charges are constructed as
\begin{eqnarray}
&~& N_{A} \equiv - i  \int d^3x~j_{A}^0
= - i \int d^3x~\left(\pi \varphi - \varphi^{\dagger} \pi^{\dagger}
+ \pi_{c_\varphi} c_{\varphi} 
 - c_{\varphi}^{\dagger} \pi_{c_{\varphi}}^{\dagger}\right),
\label{NA-non}\\
&~& N_{B} \equiv - i  \int d^3x~j_{B}^0
= - i \int d^3x~\left(\pi \varphi - \varphi^{\dagger} \pi^{\dagger}
- \pi_{c_\varphi} c_{\varphi} 
 + c_{\varphi}^{\dagger} \pi_{c_{\varphi}}^{\dagger}\right),
\label{NB-non}\\
&~& Q_{\rm F} \equiv - \int d^3x~j_{C}^{+0}
= - \int d^3x~\left(\pi c_{\varphi} 
- \varphi^{\dagger} \pi_{c_{\varphi}}^{\dagger}\right),~~
\label{QF-non}\\
&~& Q_{\rm F}^{\dagger} \equiv - \int d^3x~j_{C}^{-0}
= - \int d^3x~\left(c_{\varphi}^{\dagger} \pi^{\dagger}
- \pi_{c_{\varphi}} \varphi\right).
\label{QF-dagger-non}
\end{eqnarray}
The same algebraic relations hold as those in (\ref{QQdagger-varphi}).

The above charges are conserved and generators of global $U(1)$
and fermionic transformations for scalar fields.
They satisfy the relations,
\begin{eqnarray}
\left\{N_{A}, \phi^{\hat{a}}\right\}_{\rm PB} = 0,~~
\left\{N_{B}, \phi^{\hat{a}}\right\}_{\rm PB} = 0,~~
\left\{Q_{\rm F}, \phi^{\hat{a}}\right\}_{\rm PB} = 0,~~
\left\{Q_{\rm F}^{\dagger}, \phi^{\hat{a}}\right\}_{\rm PB} = 0,
\label{phi-PB}
\end{eqnarray}
where $\phi^{\hat{a}}$ are first class constraints (\ref{first-non}) 
and the Hamiltonian $H_{\rm M}$.
From (\ref{secondaryA-non}) -- (\ref{secondaryC--non}) and (\ref{phi-PB}), 
following relations can be considered as first class constraints,
\begin{eqnarray}
N_{A} = 0,~~ N_{B} = 0,~~ Q_{\rm F} = 0,~~ Q_{\rm F}^{\dagger} = 0.
\label{NA=0}
\end{eqnarray}

After taking the following gauge fixing conditions for the first class ones (\ref{first-non}),
\begin{eqnarray}
A^{0} = 0,~~ B^{0}  = 0,~~ {C}^{+0} = 0,~~ {C}^{-0} = 0,
\label{gf-non}
\end{eqnarray}
the system is quantized by regarding variables as operators
and imposing the same type of relations (\ref{CCR-varphi}) and (\ref{CCR-c})
on the canonical pairs.
From (\ref{NA=0}),
it is reasonable to impose the following subsidiary conditions on states,
\begin{eqnarray}
N_{A} |{\rm phys}\rangle = 0,~~ N_{B} |{\rm phys}\rangle = 0,~~
Q_{\rm F} |{\rm phys}\rangle = 0,~~
Q_{\rm F}^{\dagger} |{\rm phys}\rangle = 0.
\label{Phys-non}
\end{eqnarray}
Then, they guarantee the unitarity of our system,
though it contains negative norm states originated from
$c_{\varphi}$ and $c_{\varphi}^{\dagger}$.

\end{document}